\documentstyle[psfig,rotate,12pt]{article}
\def\M{{\cal M}}
\def\nn{\nonumber}
\begin{document} 
\title{Resonance model study of strangeness production in
$p p$ collisions \thanks{Supported in part by the Australian Research Council
and the Forschungszentrum J\"ulich}}
\author{
K. Tsushima$^1$~\thanks{ktsushim@physics.adelaide.edu.au} ,
A. Sibirtsev$^2$~\thanks{sibirt@theorie.physik.uni-giessen.de} ,
A. W. Thomas$^1$~\thanks{athomas@physics.adelaide.edu.au} \\
{\small $^1$Department of Physics and Mathematical Physics } \\
{\small and Institute for Theoretical Physics } \\
{\small University of Adelaide, SA 5005, Australia} \\
{ $^2$\small Institut f\"ur Theoretische Physik, Universit\"at Giessen} \\
{\small D-35392 Giessen, Germany} }
\date{}
\maketitle
%
%
\begin{abstract}
Results for the energy dependence of the elementary kaon production
cross sections in proton-proton collisions are reported.
Calculations are performed within an extended version 
of the resonance model which was 
used for the previous studies of elementary kaon production 
in pion-nucleon and pion-$\Delta$ 
collisions. Although the model treatment is within the {\it empirical} 
tree level (observed widths for the resonances are used), 
it is fully relativistic,
and includes all relevant baryon resonances up to 2 GeV.
One of the purposes of this study is to provide the results 
for the simulation codes of subthreshold kaon production in heavy 
ion collisions.
This is the first, 
consistent study of the elementary kaon production reactions 
including both $\pi B$ and $B B$ ($B=N, \Delta$) collisions 
on the same footing. 
Comparisons are made between the calculated results and the existing 
semi-empirical parametrizations which are 
widely used for the simulation codes,  
as well as the experimental data.
\end{abstract}

Strangeness production in heavy ion collisions is presently an issue
of intense study. The enhancement of strangeness production 
in heavy ion collisions might indicate evidence for 
a new form of nuclear matter, 
such as a quark gluon plasma~\cite{Rafelski,Nagamiya}.
Particularly, kaon production in heavy ion collisions is  
considered a promising method not only of obtaining information about 
hot, dense nuclear matter, but also of 
determining the nuclear equation of state~\cite{aic,Mosel,liko}.

For the study of kaon production in heavy ion collisions,
the elementary cross sections in 
pion-nucleon and baryon-baryon collisions are some of the
most basic and important ingredients for 
the microscopic calculation of the total kaon yield.
However, the available experimental data for the
baryon-baryon collision channels are very scarce  
(most exist for $p p$ collision channels) 
and thus could not be parameterized in a satisfactory 
way for practical use.
Indeed, there are drastic differences between the commonly used, 
semi-empirical parametrizations for (the energy
dependence of) the total kaon production
cross sections in baryon-baryon collisions
proposed by Randrup and Ko~\cite{Ko1},
and  Sch\"urmann and Zwermann~\cite{Zwermann}.
Thus, one has to parametrize these cross 
sections by relying on the theoretical 
investigations, which also describe the pion-baryon 
collision channels at the same time.

Although simulation results for kaon production in heavy
ion collisions suggest that the $N \Delta$ 
and $\Delta \Delta$ collision channels give the 
dominant contribution for the total kaon 
yield~\cite{lang,fang,hartnack},  
the present situation regarding the theoretical investigations
seems to be unsatisfactory. 
In the past, 
the reactions $p p \to N Y K$ ($Y = \Lambda, \Sigma$ hyperons)
were investigated within a  One
Boson Exchange Model (OBEM) by
Ferrari~\cite{Ferrari}, Yao~\cite{Yao}, Wu and Ko~\cite{Wu}, 
Laget~\cite{Laget} and Deloff~\cite{Deloff}. These calculations
show quite contradictory contributions from the pion and kaon exchange
processes, which can be ascribed to the different
values of the coupling constants 
and cut-off parameters applied.
Nevertheless, all of these calculations reproduce almost
perfectly {\it the small number of} available experimental data points.
This may imply the necessity of the constraints to fix 
the pion and kaon contributions consistently by investigating 
some other appropriate reactions.

Recently, Li and Ko~\cite{Ko2}, and Sibirtsev~\cite{Sibirtsev}
performed calculations of kaon production in heavy ion collisions and
proton-nucleus collisions, respectively  
by using the elementary kaon production cross sections calculated 
using the OBEM.
They adopted the results of 
the resonance model~\cite{Tsushima1,Tsushima2,Tsushima3} 
for the elementary cross sections 
$\pi B \to Y K$ in order to 
extract the $B_1 B_2 \to B_3 Y K$ cross sections.
There, an isospin average procedure is introduced
for the $\pi B \to Y K$ cross sections, and the exchanged
pion and kaon are treated on their mass shells.
Furthermore, although the kaon exchange contribution was 
included in addition to the pion exchange~\cite{Ko2,Sibirtsev}, the  
$\eta$ and $\rho$-meson exchange contributions were not studied.
To be consistent with the resonance model, 
$\eta$ and $\rho$-meson exchanges should be
included, since those resonances included 
in the model are observed to decay to the $\eta N$ 
and $\rho N$ as well as the $\pi N$ 
($\pi \Delta$) channels~\cite{Particle}.
Actually, as will be discussed later, the
$\rho$-meson exchange gives a dominant contribution 
in the present calculations.

In this article, we present the results for the energy dependence of the
total cross sections 
$p p \to N Y K$ calculated in 
the resonance model~\cite{Tsushima1,Tsushima2,Tsushima3}.
For the other reactions,  
$B_1 B_2 \to B_3 Y K$ ($B_1, B_2, B_3 = N, \Delta$), 
an extensive summary of the results of the study 
will be reported elsewhere~\cite{Tsushima4}.

The advantages of the present investigations are as follows. 
The resonance model~\cite{Tsushima1,Tsushima2,Tsushima3} 
has already been applied to the study of the 
$\pi B \to Y K$ ($B = N, \Delta$) reactions.
Although the model treatment is within the {\it empirical} tree level
(observed widths for the resonances are used), 
it is fully relativistic, and includes all relevant baryon
resonances up to 2 GeV, which are observed to decay to the 
hyperon and kaon channels.
It turned out that the model was successful in explaining the abundance of 
available experimental data for the
$\pi N \to Y K$ reactions~\cite{Landolt}. 
This means that the richness of the 
data can provide a 
sound basis to the model because the parameters 
in the model are determined so as to reproduce 
the whole available data optimally.
Thus, many of the effective interactions which are 
necessary to describe the $B_1 B_2 \to B_3 Y K$ reactions 
are fixed within our model.
Furthermore, we believe the present calculations are the first systematic
calculations for both the
$\pi B \to Y K$ and 
$B_1 B_2 \to B_3 Y K$ ($p p \to N Y K$)  reactions.

\begin{figure}[h]
{\psfig{figure=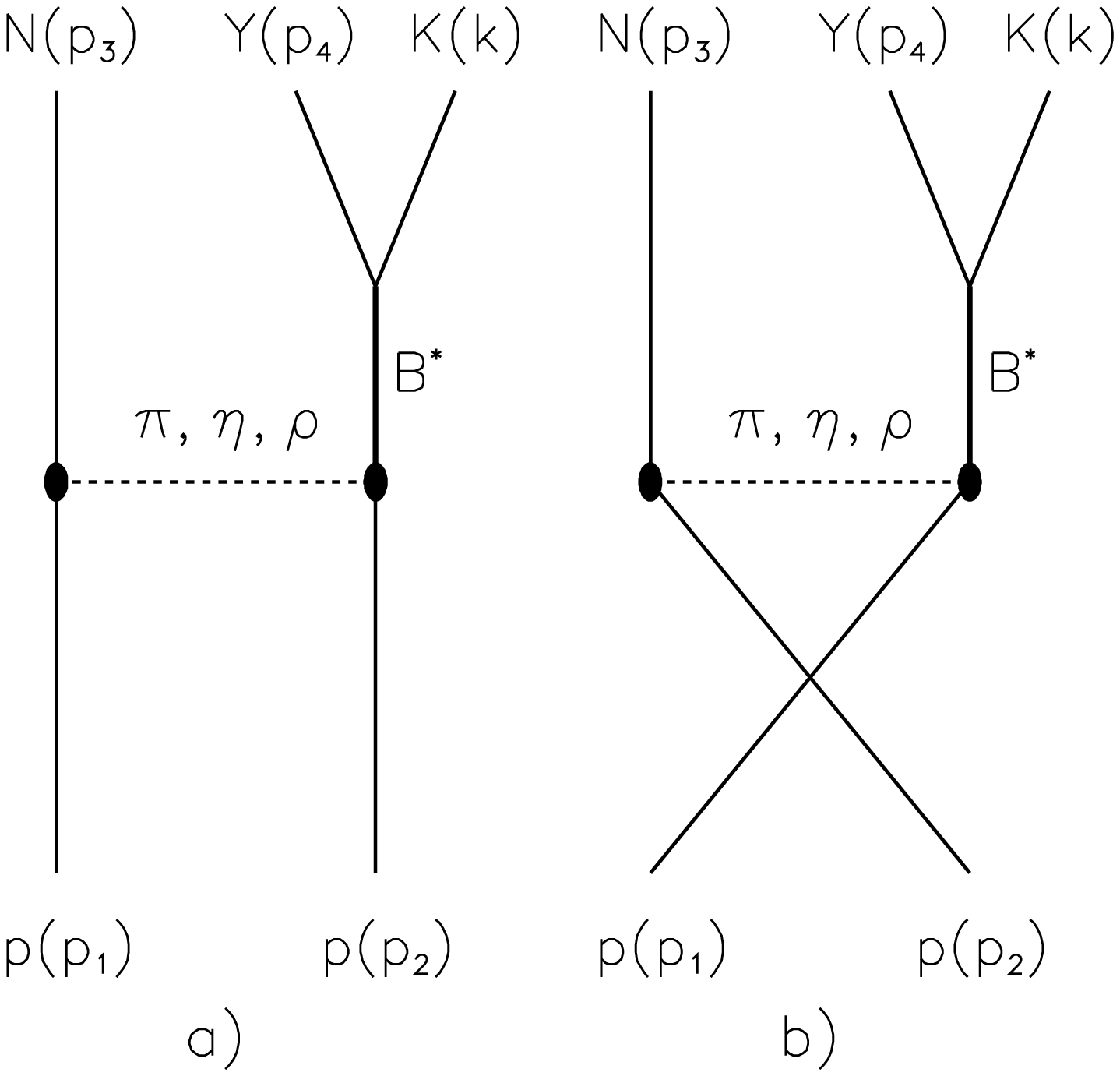,width=14cm,height=8cm}}
\caption{\label{diagram}
Processes contribute to kaon production in
$p p$ collisions. In figure, $B^*$ stands for the resonances
which are observed to decay to kaon (K) and hyperons (Y = $\Lambda, \Sigma$)
up to their masses around 2 GeV. They are,  
$N(1650) (\frac{1}{2}^-)$, $N(1710) (\frac{1}{2}^+)$, 
$N(1720) (\frac{3}{2}^+)$ and $\Delta(1920) (\frac{3}{2}^+)$.
The exchanged mesons $\pi$, $\eta$ and $\rho$ are experimentally
observed in these resonance decay channels.
}
\end{figure}

The processes relevant for the present calculations are depicted in 
Fig.~\ref{diagram}. Here $B^*$ stands for the baryon resonances which
decay to 
the kaon-hyperon channels with their masses 
up to around 2 GeV~\cite{Particle}, namely  
$N(1650) (\frac{1}{2}^-)$, $N(1710) (\frac{1}{2}^+)$,
$N(1720) (\frac{3}{2}^+)$ and $\Delta(1920) (\frac{3}{2}^+)$.
In our model, kaons are always assumed to arise from these resonances 
at the same time as the hyperons.
Once the effective Lagrangians and form factors for the vertices 
are fixed, most of the relevant coupling constants can be 
calculated by using the experimental branching ratios.
More detailed explanations of the model can be found 
in refs.~\cite{Tsushima1,Tsushima2,Tsushima3}.

Effective Lagrangian densities necessary to evaluate the
Feynman diagrams shown
in Fig.~\ref{diagram} are given by, 
\begin{eqnarray}
{\cal L}_{\pi N N} &=& 
-ig_{\pi N N} \bar{N} \gamma_5 \vec\tau N \cdot \vec\pi,
\label{pnn}\\
{\cal L}_{\pi N N(1650)} &=&  
-g_{\pi N N(1650)}
\left( \bar{N}(1650) \vec\tau N \cdot \vec\pi
+ \bar{N} \vec\tau N(1650) \cdot \vec\pi\,\, \right), 
\label{pna}\\
{\cal L}_{\pi N N(1710)} &=&  
-ig_{\pi N N(1710)}
\left( \bar{N}(1710) \gamma_5 \vec\tau N \cdot \vec\pi
+ \bar{N} \vec\tau \gamma_5 N(1710) \cdot \vec\pi\,\, \right), 
\label{pnb}\\
{\cal L}_{\pi N N(1720)} &=&
\frac{g_{\pi N N(1720)}}{m_\pi}
\left( \bar{N}^\mu(1720) \vec\tau N \cdot \partial_\mu \vec\pi
+ \bar{N} \vec\tau N^\mu(1720) \cdot \partial_\mu \vec\pi \, \right),
\label{pnc}\\
{\cal L}_{\pi N \Delta(1920)} &=& 
\frac{g_{\pi N \Delta(1920)}}{m_\pi}
\left( \bar{\Delta}^\mu(1920) \overrightarrow{\cal I} N \cdot 
\partial_\mu \vec\pi + \bar{N} {\overrightarrow{\cal I}}^\dagger 
\Delta^\mu(1920) \cdot \partial_\mu \vec\pi \, \right),
\label{pnd}\\
{\cal L}_{\eta N N} &=& 
-ig_{\eta N N} \bar{N} \gamma_5 N \eta,
\label{enn}\\
{\cal L}_{\eta N N(1710)} &=&
-ig_{\eta N N(1710)}
\left( \bar{N}(1710) \gamma_5 N \eta
+ \bar{N} \gamma_5 N(1710) \eta\,\, \right),
\label{enb}\\
{\cal L}_{\eta N N(1720)} &=&
\frac{g_{\eta N N(1720)}}{m_\eta}
\left( \bar{N}^\mu(1720) N \partial_\mu \eta
+ \bar{N} N^\mu(1720) \partial_\mu \eta \, \right),
\label{enc}\\
{\cal L}_{\rho N N} &=& - g_{\rho N N}
\left( \bar{N} \gamma^\mu \vec\tau N \cdot \vec\rho_\mu
+ \frac{\kappa}{2 m_N} \bar{N} \sigma^{\mu \nu}
\vec\tau N \cdot \partial_\mu \vec\rho_\nu \right),
\label{rnn}\\
{\cal L}_{\rho N N(1710)} &=&
-g_{\rho N N(1710)}
\left( \bar{N}(1710) \gamma^\mu \vec\tau N \cdot \vec\rho_\mu
+ \bar{N} \vec\tau \gamma^\mu N(1710) \cdot \vec\rho_\mu\,\, \right),
\label{rnb}\\
{\cal L}_{\rho N N(1720)} &=&
-ig_{\rho N N(1720)}
\left( \bar{N}^\mu(1720) \gamma_5 \vec\tau N \cdot \vec\rho_\mu
+ \bar{N} \vec\tau \gamma_5 N^\mu(1720) \cdot \vec\rho_\mu\,\, \right),
\label{rnc}\\
{\cal L}_{K \Lambda N(1650)} &=&
-g_{K \Lambda N(1650)}
\left( \bar{N}(1650) \Lambda K + \bar{K} \bar\Lambda  N(1650) \right),
\label{kla}\\
{\cal L}_{K \Sigma N(1710)} &=&
-ig_{K \Lambda N(1710)}
\left( \bar{N}(1710) \gamma_5 \Lambda K
+ \bar{K} \bar\Lambda \gamma_5 N(1710) \right),
\label{klb}\\
{\cal L}_{K \Lambda N(1720)} &=&
\frac{g_{K \Lambda N(1720)}}{m_K}
\left( \bar{N}^\mu(1720) \Lambda
\partial_\mu K + (\partial_\mu \bar{K}) \bar\Lambda N^\mu(1720) \right),
\label{klc}\\
{\cal L}_{K \Sigma N(1710)} &=&
-ig_{K \Sigma N(1710)}
\left( \bar{N}(1710) \gamma_5 \vec\tau \cdot \overrightarrow\Sigma K
+ \bar{K} \overrightarrow{\bar \Sigma} \cdot \vec\tau
\gamma_5 N(1710) \right),
\label{ksb}\\
{\cal L}_{K \Sigma N(1720)} &=&
\frac{g_{K \Sigma N(1720)}}{m_K}
\left( \bar{N}^\mu(1720) \vec\tau \cdot \overrightarrow\Sigma
\partial_\mu K + (\partial_\mu \bar{K}) \overrightarrow{\bar \Sigma}
\cdot \vec\tau N^\mu(1720) \right),
\label{ksc}\\
{\cal L}_{K \Sigma \Delta(1920)} &=&
\frac{g_{K \Sigma \Delta(1920)}}{m_K}
\left( \bar{\Delta}^\mu(1920) \overrightarrow{\cal I}
\cdot \overrightarrow\Sigma \partial_\mu K
+ (\partial_\mu \bar{K}) \overrightarrow{\bar \Sigma} \cdot
{\overrightarrow{\cal I}}^\dagger \Delta^\mu(1920) \right).
\label{ksd}
\end{eqnarray}
%
The notation and definitions appearing in the above equations are as follows: 
$\vec{\cal I}$ is the transition operator defined by
\begin{equation}
\overrightarrow{\cal I}_{M\mu} = \displaystyle{\sum_{\ell=\pm1,0}}
(1 \ell \frac{1}{2} \mu | \frac{3}{2} M)
\hat{e}^*_{\ell}, 
\end{equation}
with $M$ and $\mu$ being the third components of the 
isospin states, 
and $\vec \tau$ the Pauli matrices.
$N, N(1710), N(1720)$ and $\Delta(1920)$
stand for the fields of the nucleon
and baryon resonances. They are expressed by
$\bar{N} = \left( \bar{p}, \bar{n} \right)$,
similarly for the nucleon resonances, and
$\bar{\Delta}(1920) = ( \bar{\Delta}(1920)^{++},
\bar{\Delta}(1920)^+, \bar{\Delta}(1920)^0,
\bar{\Delta}(1920)^- )$ in isospin space.
The physical representations of the fields are, 
$K^T = \left( K^+, K^0 \right),\,\,
\bar{K} = \left( K^-, \bar{K^0} \right),\,\,
\pi^{\pm} =  (\pi_1 \mp i \pi_2)/\sqrt{2},\,
\pi^0 = \pi_3,\,\,$ similarly for the $\rho$-meson fields, and 
$\Sigma^{\pm} = (\Sigma_1 \mp i \Sigma_2)/{\sqrt{2}},\,\,
\Sigma^0 = \Sigma_3\,\,$, respectively, 
where the superscript $T$ means the 
transposition operation.
The meson fields
are defined as annihilating (creating) the physical particle
(anti-particle) states.
For the propagators $iS_F(p)$ of the spin 1/2 and
$iG^{\mu \nu}(p)$ of the spin 3/2 resonances we use:
\begin{equation}
iS_F(p) = i \frac{\gamma \cdot p + m}{p^2 - m^2 + im\Gamma^{full}}\,,
\end{equation}
\begin{equation}
iG^{\mu \nu}(p) = i \frac{-P^{\mu \nu}(p)}{p^2 - m^2 +
im\Gamma^{full}}\,,
\end{equation}
with
\begin{equation}
P^{\mu \nu}(p) = - (\gamma \cdot p + m)
\left[ g^{\mu \nu} - \frac{1}{3} \gamma^\mu \gamma^\nu
- \frac{1}{3 m}( \gamma^\mu p^\nu - \gamma^\nu p^\mu)
- \frac{2}{3 m^2} p^\mu p^\nu \right], \label{pmunu}
\end{equation}
\newline
where $m$ and $\Gamma^{full}$ stand for the mass and full decay
width of the corresponding resonances.
For the form factors $F(\vec{q})$ attached to
the meson-resonance vertices, we use 
\begin{equation}
F(\vec{q}) = 
\displaystyle{\frac{\Lambda^2}{\Lambda^2 + \vec{q}\,^2}}, 
\end{equation}
where 
$\vec{q}$ and $\Lambda$ are the three momentum of the meson and
cut-off parameter, respectively. 
The form factors, coupling constants
and cut-off parameters 
are adopted from 
refs.~\cite{Tsushima1,Tsushima2,Tsushima3}\footnote{The 
coupling constants for $g^2_{K \Sigma N(1710)}$ and 
$g^2_{K \Sigma N(1720)}$ in the tables of 
refs.~\cite{Tsushima1,Tsushima2} were listed wrongly, although 
the numerical calculations
were performed by using the correct values as given in this article.}.
For the cut-off parameters 
in the $\eta N B^*$ and $\rho N B^*$ 
vertex form factors in  $p p$ (baryon-baryon) collisions
which have not appeared in earlier work,
we use the 
same values as those for the $\pi N B^*$ vertices in order 
to reduce the number of new parameters.
The other 
coupling constants, cut-off parameters and form factors
are, as far as possible, taken from
the Bonn nucleon-nucleon potential model~\cite{Machleidt}
(Model I in TABLE B.1).
We use a dipole form factor and the tensor coupling 
constant is given by the ratio $\kappa = f_{\rho N N}/g_{\rho N N} = 6.1$ 
for the $\rho N N$ vertex.
In order to show the structure of total amplitude 
in terms of the meson exchange and resonance 
intermediate states, we explicitly write down the contributions for the 
$p p \to p \Lambda K^+$ reaction as an example:
\begin{eqnarray}
\M(p p \to p \Lambda K^+) &=& \M(\pi, N(1650)) 
+ \M(\pi, N(1710)) + \M(\pi, N(1720)) \nn \\
&+& \M(\eta, N(1710)) + \M(\eta, N(1720)) 
+ \M(\rho, N(1710)) + \M(\rho, N(1720)) \nn \\
&+& {\rm exchange}. \label{amp} 
\end{eqnarray}
On the right hand side of eq.~(\ref{amp}), the exchanged mesons
and intermediate state resonances are written inside the brackets
(c.f. Fig.~\ref{diagram}).
Since there is no way to fix adequately the
relative signs among the
amplitudes, we simply neglect all the 
interference terms (including the exchange terms).
For example, the total amplitude for 
$p p \to p \Lambda K^+$ given by eq.~(\ref{amp}) contains 
fourteen different amplitudes including the exchange amplitudes 
(diagram b) in Fig.~\ref{diagram}),
which have relative minus signs to the corresponding direct
amplitudes (diagram a) in Fig.~\ref{diagram}).
Thus there arise $2^{6}$ different 
possible relative sign combinations among them.
Concerning the ineterference effects between the direct and 
exchange amplitudes, we comment as follows.
At energy 1 GeV above the threshold,
we calculated the absolute ratios between the direct and exchange
amplitudes for the $\rho$-meson exchange (which gives
dominant contribution in our model)
in order to make a rough estimate of the interefernce effects
for the $p p \rightarrow p \Lambda K^+$ reaction. 
It turned out that the direct (exchange) process gave a
dominant contribution when the final proton was scattered in the 
forward (backward) direction.
The absolute ratios of the 
direct and exchange amplitudes were typically a few percent,
when the absolute values of the amplitudes 
became large and gave dominant contributions to the cross sections.
Furthermore, when the final proton stopped,
a sum of the absolute values of the direct and the exchange
amplitudes amounted to at most 
about a 10 percent of the forward (backward) scattering case.
Although the interference effects between the direct 
and exchange processes could be important near the threshold, the effects 
seems to be small around 1 GeV above the threshold, where the
new parameters of the model were fitted to the data. 
The effects of interference terms were
studied in refs.~\cite{Tsushima1,Tsushima2} 
for the $\pi B$ collision channels, and it turned out that 
their effects to the shape of the differential cross sections
were appreciable, but their effects to the total cross sections were
small.

For completeness, we give the relations between the $K^+$ and $K^0$ 
production cross sections treated in this article as follows:
\begin{eqnarray}
\sigma (p p \rightarrow p \Lambda K^+) 
&=& \sigma (n n \rightarrow n \Lambda K^0),\\
\sigma (n n \rightarrow n \Sigma^- K^+) 
&=& \sigma (p p \rightarrow p \Sigma^+ K^0),\\
\sigma (p p \rightarrow p \Sigma^0 K^+) 
&=& \sigma (n n \rightarrow n \Sigma^0 K^0),\\
\sigma (p p \rightarrow n \Sigma^+ K^+) 
&=& \sigma (n n \rightarrow n \Sigma^- K^0).
\end{eqnarray}

Here it should be mentioned that
the $\rho$ meson exchange turned out to 
give a dominant contribution in the present
calculations, as mentioned before.
In order to reproduce all the available data in a satisfactory manner, 
the coupling constant $g_{\rho N N}$ and 
cut-off parameter $\Lambda_\rho$ were varied. The dependence on these 
quantities is discussed below.

\begin{figure}[h]
{\psfig{figure=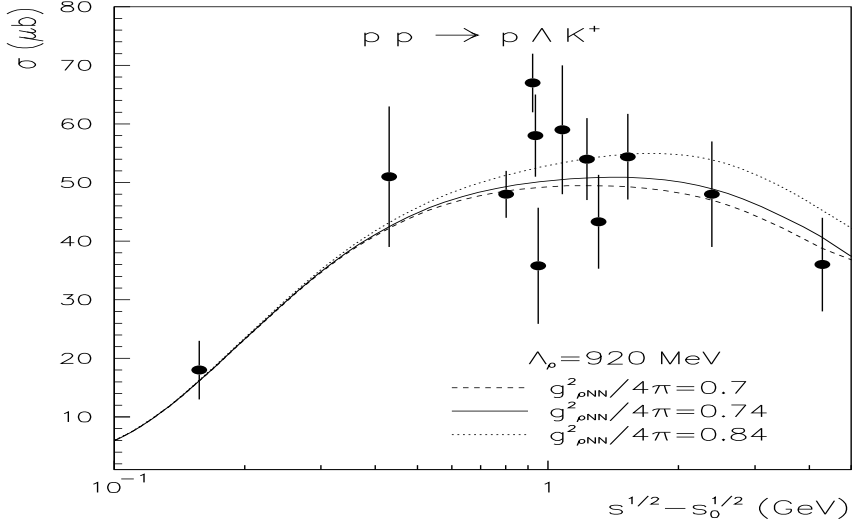,width=14cm,height=8cm}}
\caption{\label{coupling}Dependence on the coupling constant $g_{\rho N N}$
of the total cross section $pp \to p \Lambda K^+$. 
The dotted, the solid, and the dashed lines show the results obtained
with the coupling constant values $g^2_{\rho NN}/4\pi$=0.84, 
$g^2_{\rho NN}/4\pi$=0.74, and $g^2_{\rho NN}/4\pi$=0.70 
with the cut-off parameter ${\Lambda}_{\rho}=920$~MeV, respectively.
The dots show experimental data~\protect\cite{Landolt}
with error bars.}
\end{figure}

In Fig.~\ref{coupling}, the dependence on the coupling constant
$g_{\rho N N}$ of the total cross section
$p p \to p \Lambda K^+$ is shown.
Here, $s^{1/2}$ is the invariant collision energy
in the proton-proton center-of-momentum system,
and $s_0^{1/2}=m_N+m_Y+m_K$ is the threshold energy,
with $m_N$, $m_Y$ and $m_K$ being
the masses of the nucleon, hyperon and kaon, respectively.
After fixing the cut-off parameter $\Lambda_\rho$ 
to a specific value, the dependence on the coupling constant is small.

\begin{figure}[h]
{\psfig{figure=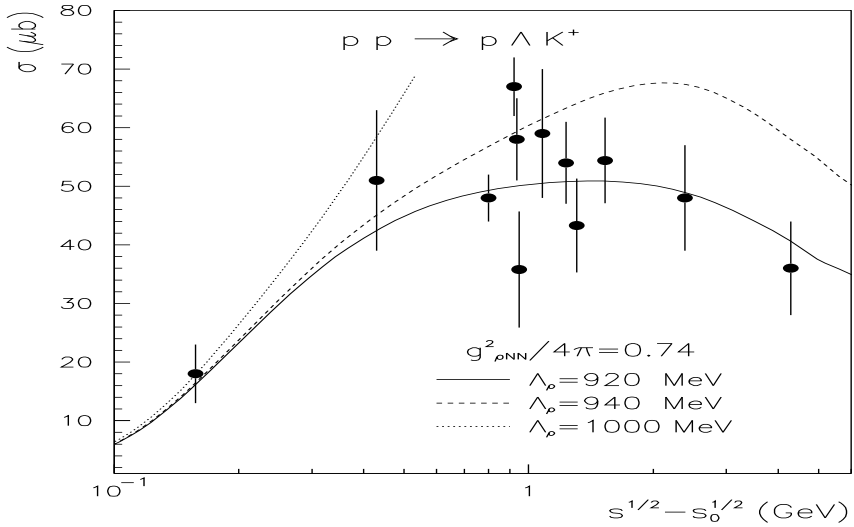,width=14cm,height=8cm}}
\caption{\label{cut}Sensitivities of the total 
cross section $pp \to p \Lambda K^+$
to the cut-off parameters $\Lambda_\rho$ 
with experimental data~\protect\cite{Landolt}. 
The dotted, the dashed and the solid lines are the 
results obtained with the
cut-off parameter values $\Lambda_\rho = $ 1000 MeV, 
$\Lambda_\rho = $ 940 MeV and $\Lambda_\rho = $ 920 MeV,
respectively. The coupling constant value 
is fixed to $g^2_{\rho NN}/4\pi = 0.74$.}
\end{figure}

Sensitivity of the 
total cross section
$p p \to p \Lambda K^+$ to the cut-off parameter values $\Lambda_\rho$ 
are shown in Fig.~\ref{cut} with the  experimental data~\cite{Landolt}.
The dependence on the cut-off parameter $\Lambda_{\rho}$ is 
rather large in the present calculation, after fixing the coupling constant 
$g_{\rho N N}$ (and $\kappa$=6.1 for the tensor coupling constant) to
adequate, {\it non-small} values. 

After all, the best values obtained to reproduce the experimental 
data are, 
$g^2_{\rho N N}/4 \pi = 0.74$ and $\Lambda_\rho = 920$ MeV.
We summarize in Table~\ref{cconst} all the relevant 
coupling constants and cut-off parameters determined and used 
for the calculations in our model.

\begin{table}
\caption{\label{cconst}
Coupling constants and cut-off parameters. 
$\kappa = f_{\rho N N}/g_{\rho N N} = 6.1$ 
for the $\rho N N$ tensor coupling is used. 
The confidence levels of the resonances listed below are, 
N(1650)****, N(1710)***, 
N(1720)**** and $\Delta$(1920)***~\protect\cite{Particle}.}
\begin{center}
\begin{tabular}{llccllc}
\hline
vertex &$g^2/4\pi$ &cut-off (MeV)& 
&vertex &$g^2/4\pi$ &cut-off (MeV)\\
\hline\\
 $\pi N N$  &$14.4$ &$1050$ & 
&$\pi N N(1650)$ &$1.12 \times 10^{-1}$ &$800$\\
 $\pi N N(1710)$ &$2.05 \times 10^{-1}$ &$800$ & 
&$\pi N N(1720)$ &$4.13 \times 10^{-3}$ &$800$\\
 $\pi N \Delta(1920)$ &$1.13 \times 10^{-1}$ &$500$ & 
&$\eta N N$ &$5.00$ &$2000$\\
 $\eta N N(1710)$ &$2.31$ &$800$ & 
&$\eta N N(1720)$ &$1.03 \times 10^{-1}$ &$800$\\
 $\rho N N$  &$0.74$ &$920$ & 
&$\rho N N(1710)$ &$3.61 \times 10^{+1}$ &$800$\\
 $\rho N N(1720)$  &$1.43 \times 10^{+2}$ &$800$ & 
&$K \Lambda N(1650)$ &$5.10 \times 10^{-2}$ &$800$\\
 $K \Lambda N(1710)$ &$3.78$ &$800$ &
&$K \Lambda N(1720)$ &$3.12 \times 10^{-1}$ &$800$\\
 $K \Sigma N(1710)$ &$4.66$ &$800$ &
&$K \Sigma N(1720)$ &$2.99 \times 10^{-1}$ &$800$\\
 $K \Sigma \Delta(1920)$ &$3.08 \times 10^{-1}$ &$500$ &
& & & \\ \\
\hline
\\
\end{tabular}
\end{center}
\end{table}

\begin{figure}[h]
{\psfig{figure=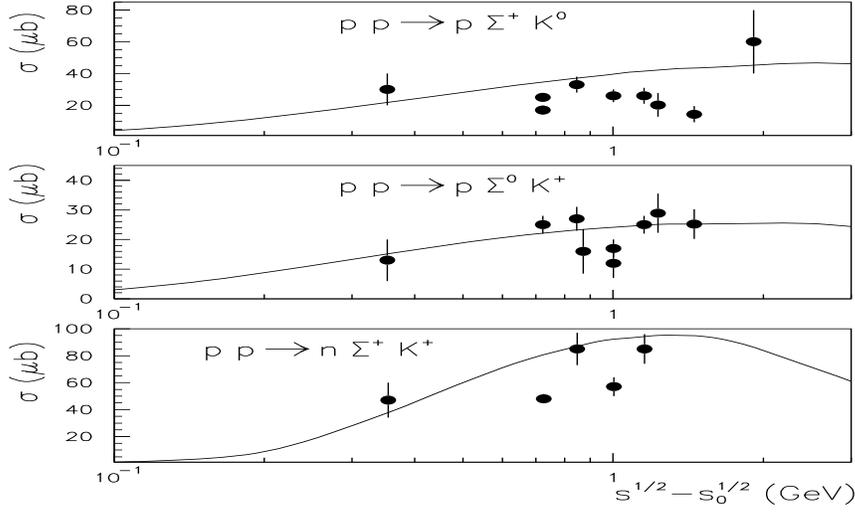,width=14cm,height=8cm}}
\caption{\label{sigma}Energy dependence of the total cross 
sections for the 
$pp \to p {\Sigma}^+ K^0$,
$pp \to p {\Sigma}^0 K^+$ and
$pp \to n {\Sigma}^+ K^+$ reactions.
The dots stand for the experimental 
data~\protect\cite{Landolt} with error bars.
The solid lines show our results.}
\end{figure}

In Fig.~\ref{sigma}  
we show the calculated energy dependence of
the total cross sections
$pp \to p {\Sigma}^+ K^0$, $pp \to p {\Sigma}^0 K^+$
and $p p \to n {\Sigma}^+ K^+$ 
in comparison with the experimental data~\cite{Landolt}.
The variations of the experimental data are rather large, but our
model can simultaneously reproduce these data well.
Remember that we do not have so much freedom in the 
number of parameters to be varied.

\begin{figure}[h]
{\psfig{figure=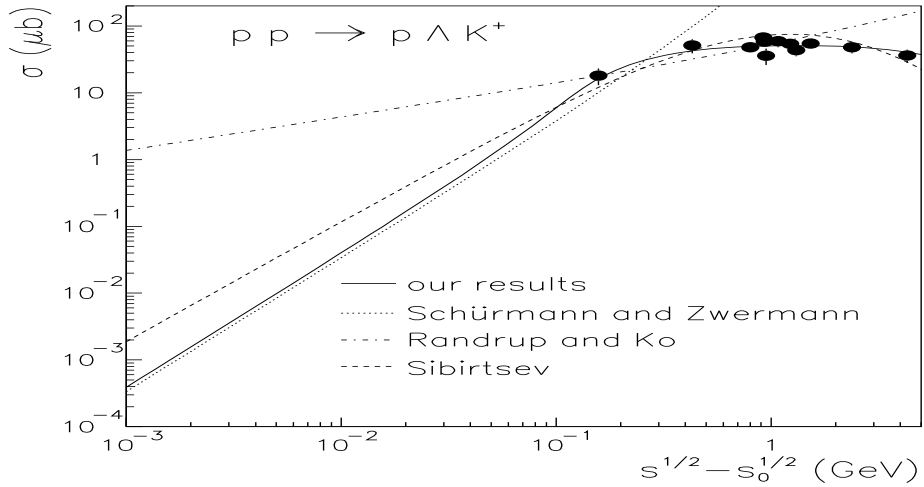,width=14cm,height=8cm}}
\caption{\label{compare}Comparisons between the existing 
semi-empirical parametrizations
and our results for the  
energy dependence of the total cross section 
$pp \to p \Lambda K^+$.
The dotted, the dashed-dotted, the dashed, 
and the solid lines stand for the parametrizations 
of Sch\"urmann and Zwermann~\protect\cite{Zwermann}, 
Randrup and Ko~\protect\cite{Ko1}, 
OBEM results of Sibirtsev~\protect\cite{Sibirtsev}, and
our results, respectively.}
\end{figure}

Now we compare our results with some of 
the existing semi-empirical parametrizations for the energy dependence of the 
total cross section $p p \to p \Lambda K^+$.
In Fig.~\ref{compare}, the dashed-dotted, the dotted,
the dashed and the solid
lines stand for the parametrizations of Randrup and Ko~\cite{Ko1},
Sch\"urmann and Zwermann~\cite{Zwermann}, 
OBEM results of Sibirtsev~\cite{Sibirtsev}, and our
results, respectively.
They illustrate quite different energy dependence and strongly disagree
when the collision energy, $s^{1/2}$, is 
close to the reaction threshold, except that the results of 
Sibirtsev~\cite{Sibirtsev} and the 
present calculations show similar
behaviour. 
We should mention here that Li and Ko \cite{Ko2} obtained 
similar results to ours by applying 
kaon and pion exchanges. There, the pion and kaon exchanges give almost 
the same contributions for the 
$p p \rightarrow p \Lambda K^+$ total cross sections, while the pion
exchange alone gives almost sufficient contribution for the
$p p \rightarrow p \Sigma K$ cross sections.
In our model, however, kaon exchange does not enter the calculation, 
since kaons are assumed to appear together with hyperons 
through the intermediate state baryon resonances.
Furthermore, the exchanged mesons in our model, $\pi, \eta$ and $\rho$ mesons
are adopted based on the observed decay channels of these baryon resonances.
However, the contribution of the $\eta$ meson exchange in our 
model is small.

We believe the strangeness production in $p p$ collisions, which 
is currently being performed at COSY-J\"ulich~\cite{TOF}, 
will provide important data to improve our understanding of 
the elementary kaon production mechanism.

In summary, we have presented new calculations for the energy dependence
of the total kaon production cross sections in proton-proton collisions
using the resonance model. The results reported in 
this article are part of the systematic 
studies for both the pion-baryon and baryon-baryon collision channels.
The investigations for {\it free space} will be complete 
in the near future~\cite{Tsushima4}. The next task to be done is to
incorporate the effects of the medium on these elementary kaon production
cross sections, which can be implimented in a unified manner based on the 
same model. Then, we hope that those in-medium modified cross sections 
can be applied to more 
realistic investigations of kaon production in heavy ion 
and proton-nucleus collisions.\\

\noindent
{\bf Acknowledgement:}\\
A. S. would like to thank W.~Cassing, C.M.~Ko and U.~Mosel for
productive discussions.\\
This work was supported in part by the Australian Research Council 
and the Forschungszentrum J\"{u}lich.
%
%
 
%
%
\newpage
%
\end{document}